\documentclass[epj,
]{svjour}
\usepackage{amsmath,amssymb,latexsym} 
\def\be{\begin{equation}}
\def\ee{\end{equation}}
\def\bea{\begin{eqnarray}}
\def\eea{\end{eqnarray}}
\def\nn{\nonumber}
\def\p{\partial}

\begin{document}

\title{Covariant anomaly and Hawking radiation from the modified black hole in the rainbow gravity theory}

\titlerunning{Covariant anomaly and Hawking radiation ...}
\authorrunning{Jun-Jin Peng \textit{et al}.}

\author{Jun-Jin Peng \and Shuang-Qing Wu \thanks{e-mail: sqwu@phy.ccnu.edu.cn}}

\institute{College of Physical Science and Technology, Central China Normal University,
Wuhan, Hubei 430079, People's Republic of China}
\date{}

\abstract{Recently, Banerjee and Kulkarni (R. Banerjee, S. Kulkarni, arXiv:0707.2449 [hep-th])
suggested that it is conceptually clean and economical to use only the covariant anomaly to derive
Hawking radiation from a black hole. Based upon this simplified formalism, we apply the covariant
anomaly cancellation method to investigate Hawking radiation from a modified Schwarzschild black
hole in the theory of rainbow gravity. Hawking temperature of the gravity's rainbow black hole is
derived from the energy-momentum flux by requiring it to cancel the covariant gravitational anomaly
at the horizon. We stress that this temperature is exactly the same as that calculated by the method
of cancelling the consistent anomaly.
\PACS{{04.70.Dy,} 04.62.+v, 11.30.-j} 
\keywords{Hawking radiation, gravity's rainbow, covariant anomaly}
}

\maketitle

\section{Introduction}

In the study of quantum theory of gravity, it is well known that the Planck length $l_p \sim 10^{-33}cm$
or its inverse, the Planck energy $E_p \sim 10^{19}GeV$ plays an important role in Quantum Gravity, and
should be taken as a universal constant, which can be thought of as the division between the classical
and the quantum description of space-time. This indicates that the value of the Planck length measured
in different inertial reference frames should remain invariant. But this feature will conflict with the
Lorentz symmetry in the special theory of relativity, because lengths are not invariant under Lorentz
transformations. Several ansatzs \cite{MSa} have been proposed to solve this paradox. One potential
candidate is the theory of doubly special relativity (DSR) \cite{MSa}, which introduces the Planck
length as a new invariant scale in the non-linear Lorentz transformation in momentum space-time. In
the theory of DSR, Lorentz invariance is only effective in the classical theory of gravity but will
be broken at the Planck scale, whereas the relativity of inertial frames and the equivalence principle
are always preserved.

When the theory of DSR was incorporated into the framework of general relativity, a rainbow gravity
scenario was proposed in \cite{MSb}. The gravity's rainbow is characterized by the feature that the
space-time geometry relies on the energy of the moving test particles in the background, which means
that different observers with probes of different energy will observe different classical geometries.
As a consequence, if a freely falling inertial observer use a system of moving particles to probe
the geometry of the space-time, the energy of the probe particles must be taken into account. That
is to say, the metric describing the space-time should contain the parameter of the probe particle's
energy $E$, forming a one-parameter energy-dependent family of metrics called as a single gravity
`rainbow' metric. As a specific example, the solution of modified Schwarzschild black holes from
gravity's rainbow has been presented in \cite{MSb}. Some interesting properties of gravity's rainbow
were investigated \cite{LLHB,LZ}, other aspects related to DSR were studied in \cite{DSR}.

On the other hand, Hawking radiation is a very intriguing phenomenon arising from the quantum effect
at the horizon. It is believed that a deeper understanding of Hawking radiation could provide some
clues to the quantum theory of gravity. During the past thirty years, many efforts have been devoted
to studying it. Recently, Robinson and Wilczek (RW) \cite{RW} proposed a new method to derive Hawking
radiation from the viewpoint of consistent anomaly cancellation. In their work, they considered a
static and spherically symmetric black hole, and treated Hawking radiation as a mechanism that cancels
a would-be gravitational anomaly at the horizon. Thus the covariance of quantum theory is protected
from potential divergences.

Since this method is very universal, it has been extended and applied to other cases
\cite{IUW,RBH,VD,XCH,BTZ,MM,SKS,JWC,WPW,BK,PW} soon. Based on the work of RW's \cite{RW}, Iso
\textit{et al}. \cite{IUW} extended it to the case of a charged black hole by considering gauge
anomaly in addition to gravitational anomaly. The case of rotating black holes was investigated
in \cite{RBH}, where the rotation acts as a induced U(1) Abelian gauge field. Subsequently considerable
efforts \cite{VD,XCH,BTZ,MM,SKS,JWC} have been done to apply the method to various kinds of space-times.
Recently, this framework was generalized to the most general case \cite{WPW} of a non-extremal black
hole with the metric determinant $\sqrt{-g} \neq 1$. However all these researches are based upon the
original RW's formalism \cite{RW,IUW} to derive Hawking radiation via the cancellation of consistent
anomalies but to fix it by requiring the regularity of covariant anomaly at the horizon. Quite recently,
Banerjee and Kulkarni \cite{BK} suggested that it is conceptually clean and economical to use only
covariant gauge and gravitational anomalies to derive Hawking radiation from charged black holes,
and the result is unaffected. While these studies successfully support the universality of the RW's
method capable of reproducing the correct Hawking temperature of a black hole, a counterexample was
presented in \cite{PW} to show that the anomaly cancellation method can only give a value of one-half
of the correct Hawking temperature of the Schwarzschild black hole in the isotropic coordinates.

With the evaporation of a black hole, its mass will continue to decrease and the Hawking temperature
will become higher and higher. At the end of Hawking radiation, physics at the Planck scale should be
taken seriously. But according to our knowledge, all the existing theory fails at the Planck scale to
describe the correct physics there. Therefore, as a kind of approximation, it may be an effective tool
to use the DSR theory to investigate physics at the Planck scale, and the study of Hawking radiation
of a modified black hole in such a theory may be helpful to shield a light on seeking a complete theory
of Quantum Gravity.

To comprehensively understand Hawking radiation and some pertinent problems, such as the information
loss paradox and the origin of black hole entropy etc, it is of great importance and significance to
investigate physics on the Plank scale, particularly the Hawking radiation from the anomaly point of
view in quantum field theory. In this paper, we'll adopt the RW's method but advised in \cite{BK} to
investigate Hawking radiation of the modified black holes from the gravity's rainbow, which stems from
the incorporation of DSR theory. Compared with the RW's original work, it should be pointed out that
Hawking flux are derived via the cancellation of covariant anomaly rather than the consistent one. In
addition, the metric considered here is energy-dependent.

The reminder of this paper begins with a brief introduction of the modified black holes from the rainbow
gravity theory in Section \ref{bhgr}. Then in Section \ref{drca}, we perform a dimension reduction and
derive the Hawking flux via the cancellation of the covariant gravitational anomaly, from which the
Hawking temperature are reproduced. The last section ends up with our conclusions.

\section{Modified Schwarzschild black holes from gravity's rainbow}
\label{bhgr}

The solution of a modified Schwarzschild black hole in the rainbow gravity theory was presented in \cite{MSb}.
A main feature of this metric is that the geometry of space-time depends on the energies of probe particles
falling in it. In other words, there are different space-time geometries for probe particles with different
energies. The rainbow metric that will be considered here takes the form \cite{MSb}
\be
 ds^2 = -\frac{1 -2M/r}{f_1^2}dt^2 +\frac{dr^2}{(1 -2M/r)f_2^2} +\frac{r^2}{f_2^2}d\Omega_2^2 \, ,
\label{metric}
\ee
where $M$ is the black-hole's mass parameter independent of the probe particle' energy $E$, $f_1$ and $f_2$
are two functions of energy $E$. It should be stressed here that the energy $E$ in the metric (\ref{metric})
is not the Arnowitt-Deser-Misner (ADM) mass of the space-time but the energy of the probes measured at
infinity by a freely falling inertial observer \cite{MSb}. Moreover, within the context of DSR, to keep
the invariance of the energy and momentum, $f_1$ and $f_2$, together with a parameter $\lambda$ of order
the Planck length, should satisfy the modified dispersion relation
\be
E^2f_1^2(E,\lambda) -p^2f_2^2(E,\lambda) = m_{0}^2 \, ,
\label{dsr}
\ee
which is a modified form of the usual energy-momentum relation. From the above equations, we see that the
metric (\ref{metric}) of the modified Schwarzschild black hole is energy-dependent, namely, dependent
on the energy of the observer, whereas the horizon $r_H = 2M$ is fixed to all observers.

Taking into account the thermodynamical properties of the modified back holes (\ref{metric}), the ADM mass
can be calculated as $M/(f_1f_2)$, which depends on the energies of the probe particles. The entropy $S =
4\pi M^2/f_2^2$ is also dependent on the probe particle's energy. This is because the quantum correction
effect on the gravity's rainbow metric is originated from the theory of DSR which has energy dependence.
The surface gravity is given by
\be
\kappa = \frac{-1}2\lim_{r\to r_H}\sqrt{\frac{-g^{rr}}{g^{tt}}}\frac{\p_rg^{tt}}{g^{tt}}
= \frac{f_2}{4Mf_1} \, ,
\ee
from which the corresponding Hawking temperature is derived
\be
T = \frac{\kappa}{2\pi}= \frac{f_2}{8\pi Mf_1} \, .
\label{temp}
\ee

Obviously, the temperature is also energy-dependent since the metric (\ref{metric}) is closely related
to the energy of particles falling in it. This means that the observers at infinity with different energy
will probe different space-time geometries with different temperatures. Therefore, the space-time (\ref{metric})
is endowed with a Plank scale modification. Furthermore, if the ADM mass is treated as the total mass of
the modified black holes, one can easily find that the mass, the temperature and the entropy obey the
differential and integral forms of the first law of black hole thermodynamics.

\section{Covariant anomaly and Hawking flux}
\label{drca}

In this section, we will perform the dimension reduction and adopt the method of covariant anomaly
cancellation to calculate the Hawking flux in the modified black hole background. Since the energy
$E$ in the metric (\ref{metric})is the one of the probes at the spatial infinity and the horizon is
independent of the energy, if an observer use a system of moving particles to probe the geometry of
the space-time, then it is assumed that the particles don't affect the physics near the horizon. In
the low order approximation, we will ignore the back effect of the radiated scalar particles on the
background space-time, but still keep the metric energy-dependent. In other words, we will regard
that the emitted  scalar particle from the black hole has no thing to do with the probe's energy,
otherwise, the analysis will be very involved.

First of all, let's reduce the dimensions of the space-time. For simplicity, we only consider the action
for a massless scalar field in the background space-time (\ref{metric})
\bea
S[\varphi] &=& -\frac{1}{2}\int d^{4}x\sqrt{-g}g^{\mu\nu}\p_{\mu}\varphi\p_{\nu}\varphi \nn \\
&=& \frac12\int dtdrd\theta d\phi sin\theta \varphi\Big\{-\frac{r^2f_1}{(1 -2M/r)f_2^3}\p_t^2 \nn \\
&& +\p_r\Big[\frac{r^2(1 -2M/r)}{f_1f_2}\p_r\Big] +\frac1{f_1f_2}\Delta_{\Omega}\Big\}\varphi \, ,
\eea
where $\Delta_{\Omega}$ is the angular Laplace operator. After performing the partial wave decomposition
in term of the spherical harmonics $\varphi = \sum_{lm}\varphi_{lm}(t, r)Y_{lm}(\theta, \phi)$, and only
keeping the dominant terms near the horizon, the action becomes
\bea
S[\varphi] &=& \frac{1}{2}\sum_{lm}\int dtdr \frac{r^2}{f_2^2}\varphi_{lm}
\Big\{-\frac{f_1}{(1 -2M/r)f_2}\p_t^2 \nn \\
&& +\p_r\Big[\frac{(1 -2M/r)f_2}{f_1}\p_r\Big]\Big\}\varphi_{lm} \, .
\eea
Therefore a free scalar field theory in the original $(3+1)$-dimensional background can be described
effectively by an infinite collection of massless fields in the $(1+1)$-dimensional space-time with
the effective metric
\be
ds^2 = -f(r)dt^2 + f^{-1}(r)dr^2 \, ,
\label{metric2}
\ee
where $f(r) = (1 -2M/r)f_2/f_1$, together with the dilaton field $\Psi = r^2/f_2^2$. Since we are
considering a static background, the contribution from the dilaton can be neglected.

Next, let's calculate the Hawking flux via the cancellation of covariant gravitational anomalies in the
two-dimensional effective space-time (\ref{metric2}). Outside the horizon, the black hole is split into
two regions: $[r_H, r_H +\epsilon]$ and $[r_H +\epsilon, +\infty)$, where the limit $\epsilon \to 0$ will
be taken ultimately. In the region $[r_H, r_H +\epsilon]$, omitting the ingoing modes that can not affect
the physics outside the horizon classically, the $(1 + 1)$-dimensional effective field theory becomes chiral
and the diffeomorphism invariance is violated at quantum level. Therefore, an additional flux is needed to
cancel the gravitational anomaly in order to keep the covariance of the underlying theory. In fact, this
current is indeed the Hawking flux. On the other hand, in the region $[r_H +\epsilon, +\infty)$, which is
far from the horizon, the energy-momentum tensor is conserved.

In \cite{RW,IUW}, the energy-momentum flux and the charge current flow are derived through the condition
that the consistent gauge and gravitational anomalies must be cancelled at the horizon, together with the
regularity requirement imposed to make the covariant energy-momentum flow and charge current vanish at the
horizon. This implies that one has to use two different anomalies, i.e. the consistent and covariant ones,
to derive the Hawking radiation. Recently, it is suggested \cite{BK} that the same work can be done by
introducing the covariant anomaly as the only one input. In the following, our analysis will be based
upon this simplified approach of cancelling the covariant anomaly.

Omitting the ingoing modes near the horizon, the effective field theory becomes chiral there and suffers
from a gravitational anomaly. The minimal form of the consistent gravitational anomaly for right handed
fields reads \cite{AWBK}
\be
\nabla_{\mu}T_{~\nu}^{\mu} = \frac{1}{96\pi\sqrt{-g}}
 \epsilon^{\beta\alpha}\p_{\alpha}\p_{\mu}\Gamma_{\nu\beta}^{\mu}
 = \frac{1}{\sqrt{-g}}\p_{\mu}N_{~\nu}^{\mu} \, .
\ee
The covariant anomaly in two dimensions, on the other hand, takes the form \cite{AWBK}
\be
\nabla_{\mu}\widetilde{T}_{~\nu}^{\mu}
= \frac{-1}{96\pi\sqrt{-g}}g_{\nu\alpha}\epsilon^{\beta\alpha}\p_{\beta}R
= \frac{1}{\sqrt{-g}}\p_{\mu}\tilde{N}_{~\nu}^{\mu} \, ,
\label{cga}
\ee
where $\epsilon^{\beta\alpha}$ is an antisymmetric tensor with convention $\epsilon^{tr} = 1$.
From the metric (\ref{metric2}), one gets
\bea
N_{~t}^r &=& \frac{1}{192\pi}\big(f^{\prime 2} +ff^{\prime\prime}\big) \nn \, , \\
\widetilde{N}_{~t}^r &=& \frac{1}{192\pi}\big(2ff^{\prime\prime} -f^{\prime 2}\big) \, .
\eea

Now introducing two scalar functions $\Theta(r) = \Theta(r -r_H -\varepsilon)$ and $H(r) = 1 -\Theta(r)$,
the total energy-momentum tensor can be written as
\be
\widetilde{T}_{~\nu}^{\mu} = \widetilde{T}_{(O)\nu}^{\mu}\Theta(r)
+\widetilde{T}_{(H)\nu}^{\mu}H(r) \, .
\ee
Here $\widetilde{T}_{(O)\nu}^{\mu}$ is covariantly conserved and $\widetilde{T}_{(H)\nu}^{\mu}$
satisfies the anomalous Eq. (\ref{cga}). Then $\widetilde{T}_{(O)t}^r$ and $\widetilde{T}_{(H)t}^r$
can be solved as
\bea
\widetilde{T}_{(O)t}^r &=& a_O \, , \\
\widetilde{T}_{(H)t}^r &=& a_H +\widetilde{N}_{~t}^r(r) -\widetilde{N}_{~t}^r(r_H) \, ,
\label{coee}
\eea
where $a_O$ and $a_H$ are two integration constants, and $a_O$ is the value of the energy flow at infinity.
In order to fix it completely, firstly we take into account Eq. (\ref{cga}) with $\nu = t$ and find
\bea
\nabla_{\mu}\widetilde{T}_{~t}^{\mu} &=& \p_r\widetilde{T}_{~t}^r
 = \p_r\big[\widetilde{N}_{~t}^rH(r)\big] \nn \\
&& +\big[\widetilde{T}_{(O)t}^r -\widetilde{T}_{(H)t}^r
 +\widetilde{N}_{~t}^r\big]\delta(r -r_H) \, .
\label{coee2}
\eea
The first term in the above equation must be cancelled by the ingoing modes to keep the theory free of
the anomaly. That is to say, there exists a Wess-Zumino term that redefines the energy-momentum tensor
$\widetilde{T}_{~t}^{\mu}$ as an anomaly-free one $\widetilde{T}_{~~t}^{\prime\mu} = \widetilde{T}_{~t}^{\mu}
-\widetilde{N}_{~t}^rH(r)$, which makes the coefficient of the delta function vanish at the horizon
\cite{BK}. Therefore, one obtains
\be
a_O = a_H-\widetilde{N}_{~t}^r(r_H) \, .
\ee
However, this condition is not sufficient to fix $a_O$ completely. One has to impose a regular boundary
condition that requires the covariant energy-momentum tensor $\widetilde{T}_{~\nu}^{\mu}$ to vanish at
the horizon, which yields $a_H = 0$. Thus the total flux of the energy-momentum tensor is
\be
a_O= -\widetilde{N}_{~t}^r(r_H) = \frac{f^{\prime 2}(r_H)}{192\pi} \, .
\label{eq14}
\ee
By comparison with the RW's original method, the energy momentum flux is determined here completely by
the covariant gravitational anomaly only. The advantage is that the concept is clean and the calculation
is very brief. On the other hand, if one adopts both of the consistent and the covariant anomalies to
derive the energy-momentum flux, he finds the result remains the same \cite{BK,PW},
\be
a_O = N_{~t}^r(r_H)= -\widetilde{N}_{~t}^r(r_H) \, .
\label{eq15}
\ee
Of course, the Hawking flux is closely related to the energy of the probe particles, because of the
energy-dependence of the effective metric.

Having got the energy-momentum flux, we now devote our attention to seeking the relation between
Hawking radiation and the gravitational anomaly in the modified black hole background. In order to
do that, let's consider the thermal flux of the black body radiation at a temperature $T$. Taken
the fermions case into account, the thermal flux is $\widetilde{T}_{~t}^r = \frac{\pi}{12}T^2$.
If we choose
\be
T = \frac{f^{\prime}(r_H)}{4\pi} = \frac{f_2}{8\pi Mf_1} \, ,
\label{eq17}
\ee
then Eq. (\ref{eq17}) coincides with Eq. (\ref{temp}), which apparently shows that the Hawking flux
is capable of cancelling the gravitational anomaly, and the temperature $T$ is just the correct Hawking
temperature of the modified black holes. Therefore our work here supports the RW's method.

\section{Conclusion}

In this paper, we have derived the correct Hawking temperature of the modified Schwarzschild black
holes in the rainbow gravity theory via the method of the cancellation of covariant anomaly. This
result supports the universality of the RW's method. It should be pointed out that the parameter
$E$ in the effective two-dimensional metric is treated as a constant in our derivation, which is
only dependent on the energy of the probe particles.

Since the temperature of the modified Schwarzschild black hole contains two energy-dependent functions
$f_1$ and $f_2$. In different energy regime, the temperature will be modified according to the concrete
expressions of these functions. In the low energy region ($E \ll E_p$), if one takes $f_1 = \exp(-E^2/2E_p^2)$ and $f_2 = 1$, then Hawking
temperature is $T = 1/(8\pi Mf_1)$. For large modified black holes with $M \gg E_p$, $E = T$ and $E^2/E_p^2
\ll 1$, then the temperature is modified as $T \simeq 1/(8\pi M) +1/(128\pi^2 M^2)$. If, however, $f_1 =
f_2 = 1$, then the temperature of the modified black hole will switch to the classical black hole case.
However, since we have neglected the back effect of the emitted particles in our analysis, therefore a
big challenge is to incorporate the correlation of the emitted particle's energy with that of the probe
particles. With this point being considered, it is possible to use DSR to address the endpoint problem
of Hawking radiation.

\section*{Acknowledgments}

S.Q.-Wu was partially supported by the Natural Science Foundation of China under Grant No. 10675051.

\end{document}